\documentclass[12pt,thmsa]{article}

\def \eq {\begin{equation}}
\def \fim-eq {\end{equation}}
\begin{document}

\author{E. S. Guerra \\
Departamento de F\'{\i}sica \\
Universidade Federal Rural do Rio de Janeiro \\
Cx. Postal 23851, 23890-000 Serop\'edica, RJ, Brazil \\
email: emerson@ufrrj.br\\
}
\title{ON THE COMPLEMENTARITY PRINCIPLE }
\maketitle

\begin{abstract}
\noindent\ We present a scheme in which we investigate the two-slit
experiment and the the principle of complementarity.\ \newline

\end{abstract}

The investigation of the double slit experiment in which we have two
micromaser cavities, each one associated with one of the slits, was proposed
before by Scully and Walther in a series of very interesting articles where
they investigate the principle of complementarity and the uncertainty
principle \cite{DSSW, Baggott, SZ} and they conclude that the principle of
complementarity is more fundamental than the uncertainty principle. \
Recently we have published an article \cite{CompPUncP1} where we have got
the same conclusion. \ In this article we proceed with a further
investigation of the subject studied in \cite{CompPUncP1}.

As in \cite{CompPUncP1} first we are going to consider a screen with two
slits $SL_{1}$ (at $\zeta _{1}$) and $SL_{2}$ \ (at $\zeta _{2}$) with two
cavities $C_{1}$ and $C_{2}$ behind respectively each slit, through which
fly Rydberg atoms of relatively long radiative lifetimes \cite{Rydat}, and
first prepared in coherent state $|\alpha \rangle _{1}$ \ and $|\alpha
\rangle _{2}$ \ \cite{SZ,Orszag} respectively and later we consider the
cavities prepared in an even coherent state $|+\rangle _{1}$ and an odd
coherent state $|-\rangle _{2}$ respectively, where 
\begin{equation}
\mid \pm \rangle _{k}=\mid \alpha \rangle _{k}\pm \mid -\alpha \rangle _{k}
\label{EOCS}
\end{equation}%
\cite{EvenOddCS}. We also assume perfect \ microwave cavities, that is, we
neglect effects due to decoherence.

Let us consider a three-level lambda atom interacting with the
electromagnetic field inside a cavity where the \ upper and the two
degenerated lower states are $|a\rangle ,$ $|b\rangle $ and $|c\rangle $
respectively, and for which the $|a\rangle \rightleftharpoons |c\rangle $
and $|a\rangle \rightleftharpoons |b\rangle $ transitions are in the far
from resonance interaction limit. The time evolution operator $U(t)$ for the
atom-field interaction in a cavity $C_{k}$ is given by \cite{Knight}%
\begin{eqnarray}
U(\tau ) &=&-e^{i\varphi a_{k}^{\dagger }a_{k}}|a\rangle \langle a|+\frac{1}{%
2}(e^{i\varphi a_{k}^{\dagger }a_{k}}+1)|b\rangle \langle b|+\frac{1}{2}%
(e^{i\varphi a_{k}^{\dagger }a_{k}}-1)|b\rangle \langle c|\ +  \nonumber \\
&&\frac{1}{2}(e^{i\varphi a_{k}^{\dagger }a_{k}}-1)|c\rangle \langle b|+%
\frac{1}{2}(e^{i\varphi a_{k}^{\dagger }a_{k}}+1)|c\rangle \langle c|,
\end{eqnarray}%
where $a_{k}$ $(a_{k}^{\dagger })$ is the annihilation (creation) operator
for the field in cavity $C_{k}$, $\varphi =2g^{2}\tau /$ $\Delta $, \ $g$ is
the coupling constant, $\Delta =\omega _{a}-\omega _{b}-\omega =\omega
_{a}-\omega _{c}-\omega $ is the detuning where \ $\omega _{a}$, $\omega
_{b} $ and $\omega _{c}$\ are the frequency of the upper and \ of the two
degenerate lower levels respectively and $\omega $ is the cavity field
frequency and $\tau $ is the atom-field interaction time. For $\varphi =\pi $%
, we get 
\begin{equation}
U(\tau )=-\exp \left( i\pi a_{k}^{\dagger }a_{k}\right) |a\rangle \langle
a|+\Pi _{k,+}|b\rangle \langle b|+\Pi _{k,-}|b\rangle \langle c|\ +\Pi
_{k,-}|c\rangle \langle b|+\Pi _{k,+}|c\rangle \langle c|,  \label{UlambdaPi}
\end{equation}%
where 
\begin{eqnarray}
\Pi _{k,+} &=&\frac{1}{2}(e^{i\pi a_{k}^{\dagger }a_{k}}+1),  \nonumber \\
\Pi _{k,-} &=&\frac{1}{2}(e^{i\pi a_{k}^{\dagger }a_{k}}-1),  \label{pi+-}
\end{eqnarray}%
and we have 
\begin{eqnarray}
\Pi _{k,+}|\alpha \rangle _{k} &=&\frac{1}{2}|+\rangle _{k},  \nonumber \\
\Pi _{k,-}|\alpha \rangle _{k} &=&\frac{1}{2}|-\rangle _{k}, \\
\Pi _{k,+}|+\rangle _{k} &=&|+\rangle _{k}, \\
\Pi _{k,+}|-\rangle _{k} &=&0,  \nonumber \\
\Pi _{k,-}|-\rangle _{k} &=&-|-\rangle _{k},  \nonumber \\
\Pi _{k,-}|+\rangle _{k} &=&0,
\end{eqnarray}%
which are easily obtained from Eqs. (\ref{pi+-}) and (\ref{EOCS}) using $%
e^{za_{k}^{\dagger }a_{k}}|\alpha \rangle _{k}=|e^{z}\alpha \rangle _{k}$ 
\cite{Louisell}.

First consider cavities $C_{1}$ and $C_{2}$ behind each slit and prepared in
coherent state $|\alpha \rangle _{1}$ \ and $|\alpha \rangle _{2}$
respectively. Before the atom $A_{1}$ passes through the slits and cavities
we have%
\begin{eqnarray}
|\Psi (t_{0})\rangle _{A_{1}-SL_{1}-SL_{2}} &=&\frac{1}{\sqrt{2}}(|\zeta
_{1}\rangle +|\zeta _{2}\rangle )|\alpha \rangle _{1}|\alpha \rangle
_{2}|b_{1}\rangle =  \nonumber \\
&=&\frac{1}{\sqrt{2}}(|\zeta _{1}\rangle +|\zeta _{2}\rangle )\frac{1}{2}%
[|+\rangle _{1}+|-\rangle _{1}]\frac{1}{2}[|+\rangle _{2}+|-\rangle
_{2}]|b_{1}\rangle .  \label{eq15}
\end{eqnarray}%
After the atom $A_{1}$ passes through the cavities $C_{1}$ and $C_{2}$ we get

\begin{eqnarray}
|\Psi (t_{0})\rangle _{A_{1}-C_{1}-C_{2}} &=&\frac{1}{4\sqrt{2}}[|\zeta
_{1}\rangle (|b_{1}\rangle |+\rangle _{1}-|c_{1}\rangle |-\rangle _{1})\
|\alpha \rangle _{2}+  \nonumber \\
&&|\zeta _{2}\rangle (|b_{1}\rangle |+\rangle _{2}-|c_{1}\rangle |-\rangle
_{2})|\alpha \rangle _{1}].  \label{eq16}
\end{eqnarray}%
Just before the atom strikes the screen at $x$, if $U(t_{1},t_{0})$ is the
time evolution operator, we have%
\begin{eqnarray}
|\Psi (t_{1})\rangle _{A_{1}-C_{1}-C_{2}} &=&\frac{1}{4\sqrt{2}}%
U(t_{1},t_{0})[|\zeta _{1}\rangle (|b_{1}\rangle |+\rangle
_{1}-|c_{1}\rangle |-\rangle _{1})\ |\alpha \rangle _{2}+  \nonumber \\
&&|\zeta _{2}\rangle (|b_{1}\rangle |+\rangle _{2}-|c_{1}\rangle |-\rangle
_{2})|\alpha \rangle _{1}],  \label{eq17}
\end{eqnarray}%
and

\begin{eqnarray}
\langle x|\Psi (t_{1})\rangle _{A_{1}-C_{1}-C_{2}} &=&\frac{1}{4\sqrt{2}}%
[\psi _{\zeta _{1}}(x,t_{1})(|b_{1}\rangle |+\rangle _{1}-|c_{1}\rangle
|-\rangle _{1})\ |\alpha \rangle _{2}+  \nonumber \\
&&\psi _{\zeta _{2}}(x,t_{1})(|b_{1}\rangle |+\rangle _{2}-|c_{1}\rangle
|-\rangle _{2})|\alpha \rangle _{1}],  \label{eq18}
\end{eqnarray}%
where%
\begin{eqnarray}
\Psi _{A_{1}}(x,t_{1}) &=&\langle x|\Psi (t_{1})\rangle
_{A_{1}-C_{1}-C_{2}}=\langle x|U(t_{1},t_{0})|\Psi (t_{0})\rangle
_{A_{1}-C_{1}-C_{2}}  \nonumber \\
\psi _{\zeta _{1}}(x,t_{1}) &=&\langle x|U(t_{1},t_{0})|\zeta _{1}\rangle ,
\\
\psi _{\zeta _{2}}(x,t_{1}) &=&\langle x|U(t_{1},t_{0})|\zeta _{2}\rangle ,
\end{eqnarray}%
where we have dropped the subindexes $C_{1}$ and $C_{2}$. The probability
density for detecting an atom at the position $x$ on the screen is

\begin{eqnarray}
\left\vert \Psi _{A_{1}}(x,t_{1})\right\vert ^{2} &=&\left[ \frac{1}{4\sqrt{2%
}}\right] ^{2}\{\left\vert \psi _{\zeta _{1}}(x,t_{1})\right\vert
^{2}+\left\vert \psi _{\zeta _{2}}(x,t_{1})\right\vert ^{2}+  \nonumber \\
&&2Re[\psi _{\zeta _{1}}^{\ast }(x,t_{1})\psi _{\zeta _{2}}(x,t_{1})[\langle
b_{1}|b_{1}\rangle _{2}\langle \alpha |+\rangle _{21}\langle +|\alpha
\rangle _{1}+  \nonumber \\
&&\langle c_{1}|c_{1}\rangle _{2}\langle \alpha |-\rangle _{21}\langle
-|\alpha \rangle _{1}]\}=  \nonumber \\
&=&\left[ \frac{1}{4\sqrt{2}}\right] ^{2}\{\left\vert \psi _{\zeta
_{1}}(x,t_{1})\right\vert ^{2}+|\psi _{\zeta _{2}}(x,t_{1})|^{2}+  \nonumber
\\
&&2Re[\psi _{\zeta _{1}}^{\ast }(x,t_{1})\psi _{\zeta
_{2}}(x,t_{1})(2-e^{-4\left\vert \alpha _{1}\right\vert
^{2}}-e^{-4\left\vert \alpha _{2}\right\vert ^{2}})]\}.
\end{eqnarray}%
If we assume $\left\vert \alpha _{1}\right\vert ^{2},\left\vert \alpha
_{2}\right\vert ^{2}\gg 1$ we have%
\begin{equation}
\left\vert \Psi _{A_{1}}(x,t_{1})\right\vert ^{2}=\left[ \frac{1}{4\sqrt{2}}%
\right] ^{2}\{\left\vert \psi _{\zeta _{1}}(x,t_{1})\right\vert
^{2}+\left\vert \psi _{\zeta _{2}}(x,t_{1})\right\vert ^{2}+4{Re}[\psi
_{\zeta _{1}}^{\ast }(x,t_{1})\psi _{\zeta _{2}}(x,t_{1})]\}.
\label{manypeaks}
\end{equation}

The injection of a coherent state $|\alpha \rangle $ in cavity $C,$ is
mathematically represented by $D(\beta )|\alpha \rangle =|\alpha +\beta
\rangle $ where $D(\beta )$ is the displacement operator $D(\beta
)=e^{(\beta a^{\dag }-\beta ^{\ast }a)}$ \cite{SZ, Orszag, Louisell}. Now,
lets us assume that after a three-level lambda atom $A_{1}$ has passed
through cavities $C_{1}$ and $C_{2}$ we inject $|-\alpha \rangle _{1}$ in
cavity $C_{1}$ and send a two-level atom $A_{2}$ resonant with the cavity,
where $|f_{2}\rangle $ and $|e_{2}\rangle $ are the lower and upper levels
respectively, through $C_{1}$. If $A_{2}$ is sent through $C_{1}$ in the
lower state, under the Jaynes-Cummings dynamics \cite{SZ,Orszag} we know
that the state $|f_{2}\rangle |0\rangle _{1}$ does not evolve, however, the
state $|f_{2}\rangle |-2\alpha \rangle _{1}$ evolves to $|e_{2}\rangle |\chi
_{e1}\rangle +|f_{2}\rangle |\chi _{f1}\rangle $, where $|\chi _{f1}\rangle
=\sum\limits_{n_{1}}C_{n_{1}}\cos (gt\sqrt{n_{1}})|n_{1}\rangle $ and $|\chi
_{e1}\rangle =-i\sum\limits_{n_{1}}C_{n_{1}+1}\sin (gt\sqrt{n_{1}+1}%
)|n_{1}\rangle $ and $C_{n_{1}}=e^{-|2\alpha _{1}|^{2}}(-2\alpha
_{1})^{n_{1}}/\sqrt{n_{1}!}$.

After we inject $|-\alpha \rangle _{1}$ in cavity $C_{1}$ we get%
\begin{eqnarray}
|\Psi (t_{0})\rangle _{A_{1}-C_{1}-C_{2}}|f_{2}\rangle  &=&\frac{1}{4\sqrt{2}%
}[|\zeta _{1}\rangle (|b_{1}\rangle (|0\rangle _{1}+|-2\alpha \rangle _{1})-
\nonumber \\
&&|c_{1}\rangle (|0\rangle _{1}-|-2\alpha \rangle _{1}))\ |\alpha \rangle
_{2}+  \nonumber \\
&&|\zeta _{2}\rangle (|b_{1}\rangle |+\rangle _{2}-|c_{1}\rangle |-\rangle
_{2})|0\rangle _{1}]|f_{2}\rangle ,  \label{EQ29}
\end{eqnarray}%
and after atom $A_{2}$ has passed through $C_{1}$ we can write the state of
the system $A_{1}-A_{2}-C_{1}-C_{2}$ as follows 
\begin{eqnarray}
|\Psi (t_{0})\rangle _{A_{1}-A_{2}-C_{1}-C_{2}} &=&\frac{1}{4\sqrt{2}}%
[|\zeta _{1}\rangle (|b_{1}\rangle (|f_{2}\rangle |0\rangle
_{1}+|e_{2}\rangle |\chi _{e1}\rangle +|f_{2}\rangle |\chi _{f1}\rangle )- 
\nonumber \\
&&|c_{1}\rangle (|f_{2}\rangle |0\rangle _{1}-|e_{2}\rangle |\chi
_{e1}\rangle -|f_{2}\rangle |\chi _{f1}\rangle ))\ |\alpha \rangle _{2}+ 
\nonumber \\
&&+|\zeta _{2}\rangle (|b_{1}\rangle |+\rangle _{2}-|c_{1}\rangle |-\rangle
_{2})|f_{2}\rangle |0\rangle _{1}].  \label{EQ30}
\end{eqnarray}%
Now, before atom $A_{1}$ strikes the screen, we detect atom $A_{2}$ in the
state $|e_{2}\rangle $. Then we get%
\begin{equation}
|\Psi (t_{0})\rangle _{A_{1}-A_{2}-C_{1}-C_{2}}=\frac{1}{\sqrt{2}}[|\zeta
_{1}\rangle (|b_{1}\rangle |\chi _{e1}\rangle -|c_{1}\rangle |\chi
_{e1}\rangle )\ |\alpha \rangle _{2}].  \label{EQ31}
\end{equation}%
Considering the evolution of atom $A_{1}$ to the screen

\begin{equation}
|\Psi (t_{1})\rangle _{A_{1}-C_{1}-C_{2}}=\frac{1}{\sqrt{2}}%
U(t_{1},t_{0})[|\zeta _{1}\rangle (|b_{1}\rangle |\chi _{e1}\rangle
-|c_{1}\rangle |\chi _{e1}\rangle )\ |\alpha \rangle _{2}],  \label{eq32}
\end{equation}%
and%
\begin{equation}
\langle x|\Psi (t_{1})\rangle _{A_{1}-C_{1}-C_{2}}=\frac{1}{\sqrt{2}}%
[\langle x|\psi _{\zeta _{1}}\rangle (|b_{1}\rangle |\chi _{e1}\rangle
-|c_{1}\rangle |\chi _{e1}\rangle )\ |\alpha \rangle _{2}],  \label{eq33}
\end{equation}%
and, dropping the subindexes $C_{1}$ and $C_{2}$,%
\begin{equation}
\left\vert \Psi _{A_{1}}(x,t_{1})\right\vert ^{2}=\left\vert \psi _{\zeta
_{1}}(x)\right\vert ^{2}.  \label{eq34}
\end{equation}%
Therefore, we conclude that atom $A_{1}$ has passed through slit $SL_{1}$
and cavity $C_{1}$ when we detect $|e_{2}\rangle $.

Now, lets us assume that just after atom $A_{1}$ has passed through cavities 
$C_{1}$ and $C_{2}$ we inject $|-\alpha \rangle _{1}$ in cavity $C_{1}$ and $%
|-\alpha \rangle _{2}$ in cavity $C_{2}$ send a two-level atom $A_{2}$
resonant with $C_{1}$, where $|f_{2}\rangle $ and $|e_{2}\rangle $ are the
lower and upper levels respectively, through $C_{1}$, and a two-level atom $%
A_{3}$ resonant with $C_{2}$, where $|f_{3}\rangle $ and $|e_{3}\rangle $
are the lower and upper levels respectively, through $C_{2}$. If $A_{2}$ is
sent through $C_{1}$ and $A_{3}$ is sent through $C_{2}$ in the lower
states, under the Jaynes-Cummings dynamics \cite{SZ, Orszag} we know that
the state $|f_{j}\rangle |0\rangle _{k}$ does not evolve, however, the state 
$|f_{j}\rangle |-2\alpha \rangle _{k}$ evolves to $|e_{j}\rangle |\chi
_{ek}\rangle +|f_{j}\rangle |\chi _{fk}\rangle $, where $|\chi _{fk}\rangle
=\sum\limits_{n_{k}}C_{n_{k}}\cos (gt\sqrt{n_{k}})|n_{k}\rangle $ and $|\chi
_{ek}\rangle =-i\sum\limits_{n_{k}}C_{n_{k}+1}\sin (gt\sqrt{n_{k}+1}%
)|n_{k}\rangle $ and $C_{n_{k}}=e^{-|2\alpha _{k}|^{2}}(-2\alpha
_{k})^{n_{k}}/\sqrt{n_{k}!}$ and where $j=2,3$ and $k=1,2.$

After we inject $|-\alpha \rangle _{1}$ in cavity $C_{1}$ and $|-\alpha
\rangle _{2}$ in cavity $C_{2}$ we get%
\begin{eqnarray}
|\Psi (t_{0})\rangle _{A_{1}-A_{2}-C_{1}-C_{2}}|f\rangle _{2}|f\rangle _{3}
&=&\frac{1}{4\sqrt{2}}[|\zeta _{1}\rangle (|b_{1}\rangle (|0\rangle
_{1}+|-2\alpha \rangle _{1})-  \nonumber \\
&&|c_{1}\rangle (|0\rangle _{1}-|-2\alpha \rangle _{1}))|0\rangle _{2}+ 
\nonumber \\
&&|\zeta _{2}\rangle (|b_{1}\rangle (|0\rangle _{2}+|-2\alpha \rangle _{2})-
\nonumber \\
&&|c_{1}\rangle (|0\rangle _{2}-|-2\alpha \rangle _{2}))|0\rangle
_{1}]|f_{2}\rangle |f_{3}\rangle ,  \nonumber \\
&&  \label{eq35}
\end{eqnarray}%
and after atom $A_{2}$ interacts with $C_{1}$ and atom $A_{3}$ interacts
with $C_{2}$ we get,

\begin{eqnarray}
|\Psi (t_{0})\rangle _{A_{1}-A_{2}-A_{3}-C_{1}-C_{2}} &=&\frac{1}{4\sqrt{2}}%
[|\zeta _{1}\rangle (|b_{1}\rangle (|f_{2}\rangle |f_{3}\rangle |0\rangle
_{1}+  \nonumber \\
&&|e_{2}\rangle |f_{3}\rangle |\chi _{e1}\rangle +|f_{2}\rangle
|f_{3}\rangle |\chi _{f1}\rangle )-  \nonumber \\
&&|c_{1}\rangle (|f_{2}\rangle |f_{3}\rangle |0\rangle _{1}-  \nonumber \\
&&|e_{2}\rangle |f_{3}\rangle |\chi _{e1}\rangle -|f_{2}\rangle
|f_{3}\rangle |\chi _{f1}\rangle ))|0\rangle _{2}+  \nonumber \\
&&|\zeta _{2}\rangle (|b_{1}\rangle (|f_{2}\rangle |f_{3}\rangle |0\rangle
_{2}+  \nonumber \\
&&|f_{2}\rangle |e_{3}\rangle |\chi _{e2}\rangle +|f_{2}\rangle
|f_{3}\rangle |\chi _{f2}\rangle )-  \nonumber \\
&&|c_{1}\rangle (|f_{2}\rangle |f_{3}\rangle |0\rangle _{2}-  \nonumber \\
&&|f_{2}\rangle |e_{3}\rangle |\chi _{e2}\rangle -|f_{2}\rangle
|f_{3}\rangle |\chi _{f2}\rangle ))|0\rangle _{1}].  \nonumber \\
&&  \label{eq36}
\end{eqnarray}%
Now, if we detect $|e_{2}\rangle $ we get%
\begin{equation}
|\Psi (t_{0})\rangle _{A_{1}-A_{3}-C_{1}-C_{2}}=\frac{1}{4\sqrt{2}}[|\zeta
_{1}\rangle (|b_{1}\rangle |f_{3}\rangle |\chi _{e1}\rangle -|c_{1}\rangle
|f_{3}\rangle |\chi _{e1}\rangle )|0\rangle _{2}],  \label{eq37}
\end{equation}%
and the only possibility is to detect $A_{3}$ in \ state $|f_{3}\rangle $
and we get 
\begin{equation}
|\Psi (t_{0})\rangle _{A_{1}-C_{1}-C_{2}}=\frac{1}{\sqrt{2}}[|\zeta
_{1}\rangle (|b_{1}\rangle |\chi _{e1}\rangle -|c_{1}\rangle |\chi
_{e1}\rangle )|0\rangle _{2}].  \label{eq38}
\end{equation}%
Making use of 
\begin{eqnarray}
\Psi _{A_{1}}(x,t_{1}) &=&\langle x|U(t_{1},t_{0})|\Psi (t_{0})\rangle
_{A_{1}-C_{1}-C_{2}},  \nonumber \\
\psi _{\zeta _{1}}(x,t_{1}) &=&\langle x|U(t_{1},t_{0})|\zeta _{1}\rangle ,
\end{eqnarray}%
we have 
\begin{equation}
\ \left\vert \Psi _{A_{1}}(x,t_{1})\right\vert ^{2}=\left\vert \psi _{\zeta
_{1}}(x,t_{1})\right\vert ^{2}  \label{onepeakxi1}
\end{equation}%
where we have dropped the subindexes $C_{1}$ and $C_{2}$.

If we now detect $|e_{3}\rangle $ we get%
\begin{equation}
|\Psi (t_{0})\rangle _{A_{1}-C_{1}-C_{2}}=\frac{1}{\sqrt{2}}[|\zeta
_{2}\rangle (|b_{1}\rangle |f_{2}\rangle |\chi _{e2}\rangle +|c_{1}\rangle
|f_{2}\rangle |\chi _{e2}\rangle )|0\rangle _{1}],  \label{eq39}
\end{equation}%
and the only possibility is to detect $A_{2}$ in \ state $|f_{2}\rangle $
and we get, 
\begin{equation}
|\Psi (t_{0})\rangle _{A_{1}-C_{1}-C_{2}}=\frac{1}{\sqrt{2}}[|\zeta
_{2}\rangle (|b_{1}\rangle |\chi _{e2}\rangle +|c_{1}\rangle |\chi
_{e2}\rangle )|0\rangle _{1}].  \label{eq40}
\end{equation}%
Making use of 
\begin{eqnarray}
\Psi _{A_{1}}(x,t_{1}) &=&\langle x|U(t_{1},t_{0})|\Psi (t_{0})\rangle
_{A_{1}-C_{1}-C_{2}},  \nonumber \\
\psi _{\zeta _{2}}(x,t_{1}) &=&\langle x|U(t_{1},t_{0})|\zeta _{2}\rangle ,
\end{eqnarray}%
we have 
\begin{equation}
\ \left\vert \Psi _{A_{1}}(x,t_{1})\right\vert ^{2}=\left\vert \psi _{\zeta
_{2}}(x,t_{1})\right\vert ^{2}  \label{onepeakxi2}
\end{equation}%
where we have dropped the subindexes $C_{1}$ and $C_{2}$. Notice that when
we detect $|e_{2}\rangle $ or $|e_{3}\rangle $ the atomic state (of $A_{2}$
or $A_{3}$) gets disentangled of the cavity field state and one of the
cavities evolves to the vacuum state ($|0\rangle _{1}$ or $|0\rangle _{2}$).
Note also that as we are injecting $|-\alpha \rangle _{1}$ in cavity $C_{1}$
and $|-\alpha \rangle _{2}$ in cavity $C_{2}$ if atom $A_{1}$ had not left
its signature ($(|\zeta _{1}\rangle +|\zeta _{2}\rangle )|\alpha \rangle _{1}
$ $|\alpha \rangle _{2}|b_{1}\rangle \rightarrow \lbrack |\zeta _{1}\rangle
(|b_{1}\rangle |+\rangle _{1}-|c_{1}\rangle |-\rangle _{1})$ $|\alpha
\rangle _{2}+|\zeta _{2}\rangle (|b_{1}\rangle |+\rangle _{2}-|c_{1}\rangle
|-\rangle _{2})|\alpha \rangle _{1}]$) in cavities $C_{1}$ and $C_{2}$ we
would never detect states $|e_{2}\rangle $ and $|e_{3}\rangle $ since atoms $%
A_{2}$ and $A_{3}$, that had been prepared in states $|f_{2}\rangle $ and $%
|f_{3}\rangle $, would interact with the vacuum fields $|0\rangle _{1}$ and $%
|0\rangle _{2}.$ It is not possible to detect $|e_{2}\rangle $ and $%
|e_{3}\rangle $ \ simultaneously which would tell us that the atom $A_{1}$,
detecting its trajectory on the classical level and behaving like a
classical particle, had passed through both slits. We can detect only $%
|e_{2}\rangle $ and $|f_{3}\rangle $ which says that the atom $A_{1}$ has
passed through slit $SL_{1}$ or $|e_{3}\rangle $ and $|f_{2}\rangle $ which
says that the atom $A_{1}$ has passed through slit $SL_{2}$ (a particle-like
behavior).

Of course until we decide to get knowledge of the localization of the atom,
everything behaves deterministically and the hole system evolves
deterministically according to the Schr\"{o}dinger equation and
entanglement, which manifest only on the quantum mechanical level, takes
place. When we decide to magnify things, that is, to come from the quantum
level to the classical level, we perform measurements using classical
apparatus and the collapse of the wave function happens and indeterminism
takes place. According to this, of course, after atom $A_{2}$ (or $A_{3}$)
has passed through cavity $C_{1}$ (or $C_{2}$) we are not sure if we are
going to detect $|e_{2}\rangle $ or $|f_{2}\rangle $ in the case of atom $%
A_{2}$. If instead of detecting $|e_{2}\rangle $ we detect $|f_{2}\rangle $
we get%
\begin{eqnarray}
|\Psi (t_{0})\rangle _{A_{1}-A_{3}-C_{1}-C_{2}} &=&\frac{1}{\sqrt{N}}[|\zeta
_{1}\rangle (|b_{1}\rangle (|f_{3}\rangle |0\rangle _{1}+|f_{3}\rangle |\chi
_{f1}\rangle )-  \nonumber \\
&&|c_{1}\rangle (|f_{3}\rangle |0\rangle _{1}-|f_{3}\rangle |\chi
_{f1}\rangle ))|0\rangle _{2}+  \nonumber \\
&&|\zeta _{2}\rangle (|b_{1}\rangle (|f_{3}\rangle |0\rangle
_{2}+|e_{3}\rangle |\chi _{e2}\rangle +|f_{3}\rangle |\chi _{f2}\rangle )- 
\nonumber \\
&&|c_{1}\rangle (|f_{3}\rangle |0\rangle _{2}-|e_{3}\rangle |\chi
_{e2}\rangle -|f_{3}\rangle |\chi _{f2}\rangle ))|0\rangle _{1}]  \nonumber
\\
&&  \label{eq41}
\end{eqnarray}%
Now, it is possible to detect atom $A_{3}$ in states $|f_{3}\rangle $ or $%
|e_{3}\rangle $. If we detect $|f_{3}\rangle $ we get%
\begin{eqnarray}
|\psi (t_{0})\rangle _{A_{1}-C_{1}-C_{2}} &=&\frac{1}{\sqrt{N}}[|\zeta
_{1}\rangle (|b_{1}\rangle (|0\rangle _{1}+|\chi _{f1}\rangle )-  \nonumber
\\
&&|c_{1}\rangle (|0\rangle _{1}-|\chi _{f1}\rangle ))|0\rangle _{2}+ 
\nonumber \\
&&|\zeta _{2}\rangle (|b_{1}\rangle (|0\rangle _{2}+|\chi _{f2}\rangle )- 
\nonumber \\
&&|c_{1}\rangle (|0\rangle _{2}-|\chi _{f2}\rangle ))|0\rangle _{1}]
\label{eq42}
\end{eqnarray}%
and we cannot decide if atom $A_{1}$ has passed through slit $SL_{1}$ or
slit $SL_{2}$.

If instead of detecting $|f_{2}\rangle $ we detect $|e_{2}\rangle $ we get
as we have shown above%
\begin{equation}
|\Psi (t_{0})\rangle _{A_{1}-A_{3}-C_{1}-C_{2}}=\frac{1}{\sqrt{2}}[|\zeta
_{1}\rangle (|b_{1}\rangle |f_{3}\rangle |\chi _{e1}\rangle -|c_{1}\rangle
|f_{3}\rangle |\chi _{e1}\rangle )|0\rangle _{2}]  \label{eq43}
\end{equation}%
and the only possibility is to detect $A_{3}$ in state $|f_{3}\rangle $ and
we can conclude that atom $A_{1}$ has passed through slit $SL_{1}$ and
cavity $C_{1}$.

Notice that if we do not send atoms $A_{2}$ and $A_{3}$ through $C_{1}$ and $%
C_{2}$ respectively, in order to get knowledge about which path the atom $%
A_{1}$ has traveled we are left with the wave function%
\begin{eqnarray}
|\Psi (t_{0})\rangle _{A_{1}-C_{1}-C_{2}} &=&\frac{1}{4\sqrt{2}}[|\zeta
_{1}\rangle (|b_{1}\rangle |+\rangle _{1}-|c_{1}\rangle |-\rangle _{1})\
|\alpha \rangle _{2}+  \nonumber \\
&&|\zeta _{2}\rangle (|b_{1}\rangle |+\rangle _{2}-|c_{1}\rangle |-\rangle
_{2})|\alpha \rangle _{1}].  \label{eq44}
\end{eqnarray}%
Then, inspecting this wave function we see that we have two contributions:
(1) atom $A_{1}$ has passed through cavity $C_{1}$ and has left cavity $%
C_{2} $ undisturbed or (2) atom $A_{1}$ has passed through cavity $C_{2}$
and has left cavity $C_{1}$ undisturbed. Therefore, we have a quantum
superposition of this two alternatives and unless we perform a classical
measurement to decide which path the atom $A_{1}$ has followed we cannot
state that it has passed through $C_{1}$ or $C_{2}$. The cavities just
stores the information that atom $A_{1}$ has passed through both the slits.
The cavities permits us to have the potential to get which-path information
in this experiment but they do not tell us which path was followed because
we are still in the quantum level and to get which-path information we have
to come to the classical level performing measurements as we have described
above sending, for instance, two two-level atoms $A_{2}$ and $A_{3}$ and
detecting their excited states by a classical apparatus. Therefore, we
cannot say that the atom $A_{1}$ has passed through just one of the slits.
It has passed through both slits since it is a quantum object represented by
a superposition of wave functions. Note that the interference fringes do not
disappear even though we have the potential possibility of getting
which-path information in the future which is stored in cavities $C_{1}$ and 
$C_{2}$. Without the cavities we get interference fringes because we do not
have acquired even the potential possibility of deciding which path has been
followed performing a classical measurement before the atom $A_{1}$ strikes
the screen. When the atom $A_{1}$ strikes the screen, we can think it shines
some light and we see in which point it has stricken the screen. Then at
this point we come from the quantum level to the classical level and we have
the collapse of the wave function. Without the cavities we say that we get
the interference pattern because the atoms behaves as waves that interfere
and we get the interference pattern, that is, we get regions on the screen
that cannot be reached by atom $A_{1}$. With the cavities the atoms are
still described by a \ wave function and behaves as a wave passing through
both slits and cavities, but we must stress again that, in this case, we do
not get interference fringes only if we perform a measurement using a
classical apparatus before atom $A_{1}$ strikes the screen to get which-path
information.

Now let us assume that cavities $C_{1}$ and $C_{2}$ are prepared in an even
coherent state $|+\rangle _{1}$ and an odd coherent state $|-\rangle _{2}$
respectively. Let us assume that we send three-level lambda atoms through
the slits and cavities. Consider an atom $A_{1}$ prepared in the state $%
|b_{1}\rangle $ flying through the double slit. Before $A_{1}$ crosses the
cavities we have%
\begin{equation}
|\Psi (t_{0})\rangle _{A_{1}-SL_{1}-SL_{2}}=\frac{1}{\sqrt{2}}(|\zeta
_{1}\rangle +|\zeta _{2}\rangle )|+\rangle _{1}|-\rangle _{2}\mid
b_{1}\rangle ,
\end{equation}%
and after it has interacted with $C_{1}$ and $C_{2},$ taking into account (%
\ref{UlambdaPi}),%
\begin{equation}
|\Psi (t_{0})\rangle _{A_{1}-C_{1}-C_{2}}=\frac{1}{\sqrt{2}}(|\zeta
_{1}\rangle \mid b_{1}\rangle -|\zeta _{2}\rangle \mid c_{1}\rangle
)|+\rangle _{1}|-\rangle _{2}.
\end{equation}%
Now, writing 
\begin{eqnarray}
\Psi _{A_{1}}(x,t_{1}) &=&\langle x|U(t_{1},t_{0})|\Psi (t_{0})\rangle
_{A_{1}-C_{1}-C_{2}},  \nonumber \\
\psi _{\zeta _{1}}(x,t_{1}) &=&\langle x|U(t_{1},t_{0})|\zeta _{1}\rangle , 
\nonumber \\
\psi _{\zeta _{2}}(x,t_{1}) &=&\langle x|U(t_{1},t_{0})|\zeta _{2}\rangle ,
\end{eqnarray}%
where we have dropped the subindexes $C_{1}$ and $C_{2}$ and where $x$ is a
point on a screen in front of the double slit screen at a certain distance $%
L $ from it, we have%
\begin{equation}
\Psi _{A_{1}}(x,t_{1})=\frac{1}{\sqrt{2}}\{\psi _{\zeta _{1}}(x,t_{1})\mid
b_{1}\rangle -\psi _{\zeta _{2}}(x,t_{1})\mid c_{1}\rangle \}.
\end{equation}%
and%
\begin{equation}
\ \left\vert \Psi _{A_{1}}(x,t_{1})\right\vert ^{2}=\frac{1}{2}[\left\vert
\psi _{\zeta _{1}}(x,t_{1})\right\vert ^{2}+\left\vert \psi _{\zeta
_{2}}(x,t_{1})\right\vert ^{2}],  \label{2peaks}
\end{equation}%
since $\langle c_{1}\mid b_{1}\rangle =0$ and if there were no cavities we
would obtain%
\begin{equation}
\ \left\vert \Psi _{A_{1}}(x,t_{1})\right\vert ^{2}=\frac{1}{2}[\left\vert
\psi _{\zeta _{1}}(x,t_{1})\right\vert ^{2}+\left\vert \psi _{\zeta
_{2}}(x,t_{1})\right\vert ^{2}+2{Re}\{\psi _{\zeta _{1}}^{\ast
}(x,t_{1})\psi _{\zeta _{2}}(x,t_{1})\}],
\end{equation}%
which presents interference fringes. Therefore, when we place cavities $%
C_{1} $ and $C_{2}$ prepared in the states $|+\rangle _{1}$ and $|-\rangle
_{2}$ respectively, the interference fringes are washed out. \ This happens
because the parity information of the cavities is transferred to the
internal state of the atom. Notice that if we detect the atomic state of $%
A_{1}$ after it has crossed the slits and before it strikes the detection
screen at $x$ and we find $\mid b_{1}\rangle ,$ we can say that the atom has
passed through slit $SL_{1}$, and if we detect $\mid c_{1}\rangle ,$ we can
say that the atom has passed through slit $SL_{2}$ and we get which-path
information (particle-like behavior) detecting the atomic state. That is,
assuming that the detection of the internal states does not disturb the
external state of motion of the atom, in the case we detect $\mid
b_{1}\rangle $ \ we get 
\begin{equation}
\left\vert \Psi _{A_{1}}(x,t_{1})\right\vert ^{2}=\left\vert \psi _{\zeta
_{1}}(x,t_{1})\right\vert ^{2},  \label{onepeakxi1+}
\end{equation}%
and in the case we detect $\mid c_{1}\rangle $ \ we get%
\begin{equation}
\left\vert \Psi _{A_{1}}(x,t_{1})\right\vert ^{2}=\left\vert \psi _{\zeta
_{2}}(x,t_{1})\right\vert ^{2}.  \label{onepeakxi2-}
\end{equation}%
Therefore, the cavities allow us to get which-path information.

Now let us consider a four-slit experiment. We are going to consider screen $%
SC_{1}$ with slits $SL_{1}$ (at $\zeta _{1})$ and $SL_{2}$ (at $\zeta _{2}$%
), a screen $SC_{2}$ with slits $SL_{3}$ (at $\eta _{1}$) and $SL_{4}$ (at $%
\eta _{2}$) distant $L_{1}=v(t_{1}-t_{0})$ from $SC_{1}$ and where $v$ is
the atom velocity, and a screen $SC_{3}$ distant $L_{2}=v(t_{2}-t_{1})$ form
\ $SC_{2}$ and where the atoms are going to strike at a point $x$. Making
use of $|\eta _{1}\rangle \langle \eta _{1}|+$ $|\eta _{2}\rangle \langle
\eta _{2}|=1$, after the atom has passed through $SL_{1}$ (and $C_{1}$) and $%
SL_{2}$ (and $C_{2}$) and before the atom passes through slits $SL_{3}$ and $%
SL_{4}$ we have%
\begin{equation}
|\Psi (t_{1})\rangle _{A_{1}-C_{1}-C_{2}}=\frac{1}{\sqrt{2}}(|\eta
_{1}\rangle \langle \eta _{1}|+|\eta _{2}\rangle \langle \eta
_{2}|)(U(t_{1},t_{0})|\zeta _{1}\rangle \mid b_{1}\rangle
-U(t_{1},t_{0})|\zeta _{2}\rangle \mid c_{1}\rangle )|+\rangle _{1}|-\rangle
_{2}
\end{equation}%
After the atom has passed through $SL_{3}$ and $SL_{4}$ we have%
\begin{eqnarray}
|\Psi (t_{1})\rangle _{A_{1}-C_{1}-C_{2}-SL_{3}-SL_{4}} &=&\frac{1}{\sqrt{2}}%
(\langle \eta _{1}|U(t_{1},t_{0})|\zeta _{1}\rangle \mid b_{1}\rangle |\eta
_{1}\rangle +\langle \eta _{2}|U(t_{1},t_{0})|\zeta _{1}\rangle \mid
b_{1}\rangle |\eta _{2}\rangle  \nonumber \\
&-&\langle \eta _{1}|U(t_{1},t_{0})|\zeta _{2}\rangle |c_{1}\rangle |\eta
_{1}\rangle -\langle \eta _{2}|U(t_{1},t_{0})|\zeta _{2}\rangle \mid
c_{1}\rangle |\eta _{2}\rangle )|+\rangle _{1}|-\rangle _{2}
\end{eqnarray}%
Now writing%
\begin{eqnarray}
\psi _{\zeta _{1}}(\eta _{1},t_{1}) &=&\langle \eta
_{1}|U(t_{1},t_{0})|\zeta _{1}\rangle ,  \nonumber \\
\psi _{\zeta _{1}}(\eta _{2},t_{1}) &=&\langle \eta
_{2}|U(t_{1},t_{0})|\zeta _{1}\rangle ,  \nonumber \\
\psi _{\zeta _{2}}(\eta _{1},t_{1}) &=&\langle \eta
_{1}|U(t_{1},t_{0})|\zeta _{2}\rangle ,  \nonumber \\
\psi _{\zeta _{2}}(\eta _{2},t_{1}) &=&\langle \eta
_{2}|U(t_{1},t_{0})|\zeta _{2}\rangle ,
\end{eqnarray}%
we have%
\begin{eqnarray}
|\Psi (t_{1})\rangle _{A_{1}-C_{1}-C_{2}-SL_{3}-SL_{4}} &=&\frac{1}{\sqrt{2}}%
(\psi _{\zeta _{1}}(\eta _{1},t_{1})|\eta _{1}\rangle \mid b_{1}\rangle
+\psi _{\zeta _{1}}(\eta _{2},t_{1})|\eta _{2}\rangle \mid b_{1}\rangle 
\nonumber \\
&-&\psi _{\zeta _{2}}(\eta _{1},t_{1})|\eta _{1}\rangle |c_{1}\rangle -\psi
_{\zeta _{2}}(\eta _{2},t_{1})|\eta _{2}\rangle \mid c_{1}\rangle )|+\rangle
_{1}|-\rangle _{2}
\end{eqnarray}%
We now let the atom evolve toward screen $SC_{3}$ and be detected at $x$ and
we have%
\begin{eqnarray}
\langle x|\Psi (t_{2})\rangle _{A_{1}-C_{1}-C_{2}-SL_{3}-SL_{4}} &=&\frac{1}{%
\sqrt{2}}(\psi _{\zeta _{1}}(\eta _{1},t_{1})\langle x|U(t_{2},t_{1})|\eta
_{1}\rangle \mid b_{1}\rangle +\psi _{\zeta _{1}}(\eta _{2},t_{1})\langle
x|U(t_{2},t_{1})|\eta _{2}\rangle \mid b_{1}\rangle  \nonumber \\
&-&\psi _{\zeta _{2}}(\eta _{1},t_{1})\langle x|U(t_{2},t_{1})|\eta
_{1}\rangle |c_{1}\rangle -\psi _{\zeta _{2}}(\eta _{2},t_{1})\langle
x|U(t_{2},t_{1})|\eta _{2}\rangle \mid c_{1}\rangle )|+\rangle _{1}|-\rangle
_{2}  \nonumber \\
&&
\end{eqnarray}%
and we write%
\begin{eqnarray}
\psi _{\eta _{1}}(x,t_{2}) &=&\langle x|U(t_{2},t_{1})|\eta _{1}\rangle , 
\nonumber \\
\psi _{\eta _{2}}(x,t_{2}) &=&\langle x|U(t_{2},t_{1})|\eta _{2}\rangle , 
\nonumber \\
\psi _{\eta _{1}}(x,t_{2}) &=&\langle x|U(t_{2},t_{1})|\eta _{1}\rangle , 
\nonumber \\
\psi _{\eta _{2}}(x,t_{2}) &=&\langle x|U(t_{2},t_{1})|\eta _{2}\rangle ,
\end{eqnarray}%
and we have%
\begin{eqnarray}
\langle x|\Psi (t_{2})\rangle _{A_{1}-C_{1}-C_{2}-SL_{3}-SL_{4}} &=&\frac{1}{%
\sqrt{2}}(\psi _{\zeta _{1}}(\eta _{1},t_{1})\psi _{\eta _{1}}(x,t_{2})\mid
b_{1}\rangle +\psi _{\zeta _{1}}(\eta _{2},t_{1})\psi _{\eta
_{2}}(x,t_{2})\mid b_{1}\rangle  \nonumber \\
&-&\psi _{\zeta _{2}}(\eta _{1},t_{1})\psi _{\eta
_{1}}(x,t_{2})|c_{1}\rangle -\psi _{\zeta _{2}}(\eta _{2},t_{1})\psi _{\eta
_{2}}(x,t_{2})\mid c_{1}\rangle )|+\rangle _{1}|-\rangle _{2}  \nonumber \\
&&
\end{eqnarray}%
Dropping the subindexes $C_{1},C_{2},SL_{3}$ and $SL_{4}$ we finally have
the probability of the atom be detected at $x$%
\begin{eqnarray}
\left\vert \Psi _{A_{1}}(x,t_{2})\right\vert ^{2} &=&\frac{1}{2}\{\left\vert
\psi _{\zeta _{1}}(\eta _{1},t_{1})\right\vert ^{2}\left\vert \psi _{\eta
_{1}}(x,t_{2})\right\vert ^{2}+\left\vert \psi _{\zeta _{1}}(\eta
_{2},t_{1})\right\vert ^{2}\left\vert \psi _{\eta _{2}}(x,t_{2})\right\vert
^{2}+  \nonumber \\
&&\left\vert \psi _{\zeta _{2}}(\eta _{1},t_{1})\right\vert ^{2}\left\vert
\psi _{\eta _{1}}(x,t_{2})\right\vert ^{2}+\left\vert \psi _{\zeta
_{2}}(\eta _{2},t_{1})\right\vert ^{2}\left\vert \psi _{\eta
_{2}}(x,t_{2})\right\vert ^{2}+  \nonumber \\
&&2{Re}[(\psi _{\zeta _{1}}^{\ast }(\eta _{1},t_{1})\psi _{\zeta _{1}}(\eta
_{2},t_{1})+\psi _{\zeta _{2}}^{\ast }(\eta _{1},t_{1})\psi _{\zeta
_{2}}(\eta _{2},t_{1}))\psi _{\eta _{1}}^{\ast }(x,t_{2})\psi _{\eta
_{2}}(x,t_{2})]\}  \label{psiA1t2}
\end{eqnarray}%
From the above expression we see that if we have two sharp peaks centered on 
$\eta _{1}$ and $\eta _{2}$, that is, if $\psi _{\zeta _{1}}(\eta
_{2},t_{1})=\psi _{\zeta _{2}}(\eta _{1},t_{1})=0$ and $\psi _{\zeta
_{1}}(\eta _{1},t_{1})\neq 0$ and \ $\psi _{\zeta _{2}}(\eta _{2},t_{1})\neq
0$ we get%
\begin{equation}
\left\vert \Psi _{A_{1}}(x,t_{2})\right\vert ^{2}=\frac{1}{2}\{\left\vert
\psi _{\zeta _{1}}(\eta _{1},t_{1})\right\vert ^{2}\left\vert \psi _{\eta
_{1}}(x,t_{2})\right\vert ^{2}+\left\vert \psi _{\zeta _{2}}(\eta
_{2},t_{1})\right\vert ^{2}\left\vert \psi _{\eta _{2}}(x,t_{2})\right\vert
^{2}\}
\end{equation}%
and we have no interference. However, if $\psi _{\zeta _{1}}(\eta
_{2},t_{1})\neq 0$ and $\psi _{\zeta _{2}}(\eta _{1},t_{1})\neq 0$ the
interference term in (\ref{psiA1t2}) will not vanish.

Concluding, we can have a setup in which the mere introduction of a two-slit
screen on the way of atom $A_{1}$ to the detection screen makes the
observable behavior of atoms quite different of classical particles even in
the situation in which the atoms were going to present a particle-like \
behavior. Of course Quantum Mechanics is the fundamental theory and all
particles must behave according to it. Considering that the momentum of a
particle is given by $p=h/\lambda $ we see that for classical particles, as
the mass $m$ \ is large, the momentum $p=mv$ is also large. Due to smallness
of the Planck constant \ $h$ and the largeness \ of the momentum $p$, the
wavelength $\lambda $ will be extremely small and this is why we do not
observe interference for classical particles. More importantly, we notice
that in Ref. \cite{SZ} it is stated: "...The mere fact that we could in
principle have which-path information is enough to rub out the fringes..."
This is true for the systems studied in \cite{DSSW, SZ, CompPUncP1}. In
these references, the mere presence of microwave cavities behind the slits
wash out the interference fringes even though we do not make any\ direct
measurement to get which-path information. However, the previous statement a
strong one and is not always true as we have seen in the present article. We
have seen above that if the cavities are prepared in a coherent state, we
can get which-path information but the interference fringes do not vanish
due to only the presence of the cavities behind the slits. The fringes
disappear only when we perform a measurement, after the atoms had passed
through the cavities, and we get which path information. Therefore, we
stress that the potential to get which-path information does not always
destroys the interference fringes as we see in the case studied in the
present article in which the cavities are prepared in a coherent state.


\begin{thebibliography}{9}
\bibitem{DSSW} M. O. Scully and H. Walther, Phys. Rev. A \textbf{39}, 5229
(1989); M. O. Scully, B-G Englert and H. Walther, Nature \textbf{351}, 111
(1991); M. O. Scully and H. Walther, Phys. Rev. A \textbf{49}, 1562 (1994);
M. O. Scully and H. Walther, Found. Phys. \textbf{28}, 399 (1998).

\bibitem{Baggott} J. Baggott, New Sci. \textbf{133}, 36 (1992).

\bibitem{SZ} M. O. Scully and M. S. Zubairy, \ Quantum Optics, Cambridge
Univ. Press, Cambridge, 1997.

\bibitem{CompPUncP1} E. S. Guerra, Opt. Commun. \textbf{234}, 295 (2004).

\bibitem{Rydat} A. A. Radzig and B. M. Smirnov, \textit{Reference Data on
Atoms, Molecules, and Ions} (Springer-Verlag, Berlin, 1985); T. F.
Gallagher, \textit{Rydberg Atoms }(Cambridge Univ. Press, Cambridge, 1984).

\bibitem{Orszag} M. Orszag, Quantum Optics, Springer-Verlag, Berlin, 2000.

\bibitem{EvenOddCS} E. S. Guerra, B. M. Garraway and P. L. Knight, Phys.
Rev. A \textbf{55}, 3482 (1997).

\bibitem{Knight} P. L. Knight, Phys. Scr. \textbf{T12}, 51 (1986); S. J. D.
Phoenix and P. L. Knight, J. Opt. Soc. Am. B \textbf{7}, 116 (1990).

\bibitem{Louisell} W. H. Louisell, \textit{Quantum Statistical Properties of
Radiation}, (Wiley,\textbf{\ }New York, 1973).
\end{thebibliography}
\end{document}